\documentclass[%
reprint,
superscriptaddress,
 amsmath,amssymb,
 aps
]{revtex4-2}

\usepackage{graphicx}%
\usepackage{multirow}%
\usepackage{amsmath,amssymb,amsfonts}%
\usepackage{amsthm}%
\usepackage{mathrsfs}%
\usepackage[title]{appendix}%
\usepackage{xcolor}%
\usepackage{textcomp}%
\usepackage{booktabs}%
\usepackage{listings}%
\usepackage{indentfirst}
\usepackage{float}
\usepackage{hyperref}
\usepackage[all]{hypcap}
\hypersetup{
    colorlinks=true,
    linkcolor=blue,
    filecolor=blue,      
    urlcolor=blue,
    citecolor=blue
    }

\usepackage{lineno} 
\usepackage{colortbl}
\makeatletter
    \def\CT@@do@color{%
      \global\let\CT@do@color\relax
            \@tempdima\wd\z@
            \advance\@tempdima\@tempdimb
            \advance\@tempdima\@tempdimc
    \advance\@tempdimb\tabcolsep
    \advance\@tempdimc\tabcolsep
    \advance\@tempdima2\tabcolsep
            \kern-\@tempdimb
            \leaders\vrule
                    \hskip\@tempdima\@plus  1fill
            \kern-\@tempdimc
            \hskip-\wd\z@ \@plus -1fill }
    \makeatother

\usepackage{soul} 
\usepackage{bm}

\newcommand{\beq}{\begin{equation}}
\newcommand{\eeq}{\end{equation}}
\newcommand{\beqa}{\begin{eqnarray}}
\newcommand{\eeqa}{\end{eqnarray}}

\usepackage{titlesec}
\titleformat*{\section}{\centering\footnotesize\bfseries\uppercase}

\usepackage[paperwidth=210mm,paperheight=297mm,centering,hmargin=1.7cm,vmargin=1.5cm]{geometry}

\newcommand{\be}{\begin{equation}}
\newcommand{\ee}{\end{equation}}
\newcommand{\bea}{\begin{eqnarray}}
\newcommand{\eea}{\end{eqnarray}}

\begin{document}

\title{Universal Random Matrix Behavior of a Fermionic Quantum Gas}

\author{Maxime Dixmerias}
\thanks{These authors contributed equally to this work.}
\affiliation{Laboratoire Kastler Brossel, ENS-Universit\'{e} PSL, CNRS, Sorbonne Universit\'{e}, Coll\`{e}ge de France, 24 rue Lhomond, 75005, Paris, France}
\author{Giuseppe Del Vecchio Del Vecchio}
\thanks{These authors contributed equally to this work.}
\affiliation{LPTMS, CNRS, Universit\'e Paris-Sud, Universit\'e Paris-Saclay, 91405 Orsay, France}
\author{Cyprien Daix}
\affiliation{Laboratoire Kastler Brossel, ENS-Universit\'{e} PSL, CNRS, Sorbonne Universit\'{e}, Coll\`{e}ge de France, 24 rue Lhomond, 75005, Paris, France}
\author{Joris Verstraten}
\affiliation{Laboratoire Kastler Brossel, ENS-Universit\'{e} PSL, CNRS, Sorbonne Universit\'{e}, Coll\`{e}ge de France, 24 rue Lhomond, 75005, Paris, France}
\author{Tim de Jongh}
\thanks{Present address: JILA, National Institute of Standards and Technology, and Department of Physics, University of Colorado, Boulder, CO 80309, USA}
\affiliation{Laboratoire Kastler Brossel, ENS-Universit\'{e} PSL, CNRS, Sorbonne Universit\'{e}, Coll\`{e}ge de France, 24 rue Lhomond, 75005, Paris, France}
\author{Bruno Peaudecerf}
\affiliation{Laboratoire Collisions Agr\'egats R\'eactivit\'e, UMR 5589, FERMI, UT3, Universit\'e de Toulouse, CNRS, 118 Route de Narbonne, 31062, Toulouse CEDEX 09, France}
\author{Pierre Le Doussal}
\affiliation{Laboratoire de Physique de l'\'Ecole Normale Sup\'erieure, CNRS, ENS $\&$ PSL University, Sorbonne Universit\'e, Universit\'e Paris Cit\'e, 75005 Paris, France}
\author{Gr\'egory Schehr}
\affiliation{Sorbonne Universit\'e, Laboratoire de Physique Th\'eorique et Hautes Energies, CNRS UMR 7589, 4 Place Jussieu, 75252 Paris Cedex 05, France}
\author{Tarik Yefsah}
\thanks{Correspondence to be addressed to: \href{mailto:tarik.yefsah@lkb.ens.fr}{tarik.yefsah@lkb.ens.fr}}
\affiliation{Laboratoire Kastler Brossel, ENS-Universit\'{e} PSL, CNRS, Sorbonne Universit\'{e}, Coll\`{e}ge de France, 24 rue Lhomond, 75005, Paris, France}

\date{\today}

\begin{abstract}
\quad The pursuit of universal governing principles is a foundational endeavor in physics, driving breakthroughs from thermodynamics to general relativity and quantum mechanics. In 1951, Wigner introduced the concept of a statistical description of energy levels of heavy atoms \cite{wigner1951}, which led to the rise of Random Matrix Theory (RMT) in physics \cite{dyson1962,dyson1962a,mehta1991,forrester2010}. The theory successfully captured spectral properties across a wide range of atomic systems, circumventing the complexities of quantum many-body interactions. Rooted in the fundamental principles of stochasticity and symmetry, RMT has since found applications and revealed universal laws in diverse physical contexts, from quantum field theory to disordered systems and wireless communications \cite{akemann2015}. A particularly compelling application arises in describing the mathematical structure of the many-body wavefunction of non-interacting Fermi gases, which underpins a complex spatial organization driven by Pauli{'}s exclusion principle \cite{castin2007,dean2016,holten2021}.
However, experimental validation of the counting statistics predicted in such systems has remained elusive. Here, we probe at the single-atom level ultracold atomic Fermi gases made of two interacting spin states, obtaining direct access to their counting statistics in situ. Our measurements show that, while the system is strongly attractive, each spin-component is extremely well described by RMT predictions based on Fredholm determinants. Our results constitutes the first experimental validation of the Fermi-sphere point process \cite{torquato2018,dean2019} through the lens of RMT, and establishes its relevance for strongly-interacting systems. 
\end{abstract}

\maketitle

A point process is a mathematical framework for modeling random collections of points in time or space. Each ``point" corresponds to an event, such as the location of a star in the sky, the firing of a neuron, or the arrival of a bus. Unlike single random variables, point processes describe entire configurations of events. The most fundamental example is the Poisson point process, where events occur independently and uniformly, often serving as a model of complete randomness. More complex processes display clustering or repulsion between points, reflecting correlations that arise in various systems. Point processes are central in fields ranging from spatial analysis and telecommunications to neuroscience and quantum physics, providing powerful tools for understanding the structure and dynamics of random events across disciplines.

In a non-interacting Fermi gas, the underlying point process (governing the probability of finding particles at given spatial coordinates) is driven by the Pauli exclusion principle. 
It is a paradigmatic example of a determinantal point process, for which all correlation functions can be expressed as determinants of an elementary kernel function~\cite{macchi1975,johansson2005,borodin2009,forrester2010}. 
In two dimensions (2D), rotational symmetry allows to relate the Fermi point process with the complex Wishart-Laguerre ensemble \cite{forrester2010}, a class of random matrices with applications in quantum transport \cite{beenakker1997}, finance \cite{potters2020}, multivariate statistics \cite{wishart1928}, and wireless communications \cite{moustakas2017}. This mapping contrasts with the Gaussian ensembles typically associated with one-dimensional systems, and highlights the broader relevance of RMT in higher-dimensional quantum gases \cite{torquato2018,dean2019,smith2021,gouraud2022}.

Using recently developed continuum quantum gas microscopy \cite{verstraten2025,dejongh2025,yao2025,xiang2025}, we probe at the single-atom level a degenerate 2D Fermi gas composed of two spin states with strong attractive interactions. Previous studies of counting statistics in quantum gases, which concern bosonic systems, probed the momentum~\cite{perrier2019,herce2023,allemand2025}, time~\cite{ottl2005}, or energy domain~\cite{frisch2014}. In this work we access the full counting statistics of each spin-component in situ, measuring the probability of finding a given number $N$ of atoms within circular probe regions of space with a resolution well below the interparticle spacing. We compare the data to RMT calculations based on Fredholm determinants. This required extending determinantal point process techniques to finite temperatures, and pushing the precision of calculations to the level required by experimental data. The experimentally measured statistics agree strikingly well with RMT theoretical predictions for the ideal gas, at both near-zero and finite temperature, and without any fitting parameters.

\begin{figure*}[!t]
    \centering
    \includegraphics[width=\textwidth]{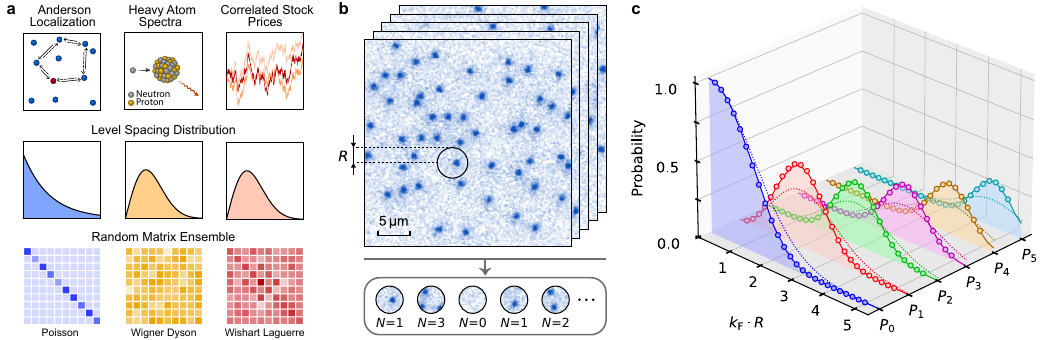}
    \caption{\textbf{Same-Spin Counting Statistics in an Interacting Fermi Gas.} \textbf{(a)} Random matrix theory (RMT) allows to describe universal statistical behaviors in complex systems, from the spectra of localized systems or atomic nuclei to financial trends, by means of several random matrix ensembles. \textbf{(b)} Continuum gas microscopy images give direct access to the number of fermions in a disk of radius $R$. \textbf{(c)} Measured probability to find $N$ atoms ($P_N$, $0\leq N \leq 5$) in a disk of variable adimensional radius $k_{\rm F}R$ (circles). The measurement is performed on one spin component of an attractive two-component Fermi gas of reduced temperature $T/T_{\rm F}=0.15(1)$. The standard deviation of the data, obtained from an average over 100 disk locations, is smaller than the marker size. Solid lines show the theoretical prediction from Fredholm determinant calculations at the experimentally measured temperature. Dotted lines correspond to Poisson statistics, describing a classical (uncorrelated) gas.}
    \label{fig:fig1}
\end{figure*}

These observations are not expected a priori: the system is well in the interacting regime, yet the measured spatial organization follows the repulsive structure of eigenvalues of Wishart-Laguerre random matrices, which maps to the ideal Fermi gas. Beyond number statistics, we measure the distribution of nearest-neighbor distances, providing a direct observation of the spacing statistics of a 2D quantum gas. These measurements reveal a smooth crossover from a strongly correlated regime at low temperature (governed by the Pauli exclusion principle) to a classical, Poissonian regime at high temperature, where correlations vanish. Our results represent a unique experimental measurement of the Fermi-sphere point process \cite{torquato2018,dean2019}, and demonstrate the unexpected relevance of RMT for the study of strongly interacting quantum systems. 

Our experiment begins with a two-component spin mixture of $^6$Li atoms, prepared with equal spin populations and confined to a single plane by a laser-induced trap that provides strong confinement along the vertical $z$-direction. In the $xy$-plane, the atoms experience a nearly harmonic potential with slow spatial variations, allowing the use of the local density approximation to extract properties of the homogeneous system \cite{castin2007}. At the ultracold temperatures relevant here, interactions occur exclusively via $s$-wave collisions, which are forbidden between identical fermions due to the Pauli exclusion principle, and thus only arise between atoms in different spin states. These inter-spin interactions are attractive and tunable via a magnetic field near a Feshbach resonance located at 690\,G.

The interaction strength is characterized by the dimensionless parameter $\eta = \log(k_{\rm F}a)$, where $k_{\rm F} = \sqrt{4\pi n}$ is the Fermi wavevector (with $n$ the density of one spin-component), and $a$ the two-dimensional scattering length (see Methods). The weakly attractive regime, $\eta \gg 1$, corresponds to a BCS superfluid at low temperature. In this work, we study various samples, with interactions ranging from extremely weak to relatively strong attraction ($\eta \approx 20$ to $\eta \approx 2$), and reduced temperatures ranging from $T/T_{\rm F} \approx 0.1$ to $20$, where $T_{\rm F} = 2 \pi n \hbar^2 / m k_{\rm B}$ is the Fermi temperature and $m$ the atomic mass (see Refs. \cite{dixmerias2025} and \cite{daix2025} for details on the preparation).

We probe the system using continuum quantum gas microscopy, which allows for single-atom-resolved imaging in continuous space. This technique involves freezing the atomic motion by rapidly switching on a deep optical lattice and then illuminating the atoms with cooling light to induce fluorescence while holding them in individual lattice sites \cite{verstraten2025,dejongh2025}. Before applying the lattice, we remove one spin component, enabling measurement of the spatial distribution of a single spin species. A typical experimental image is shown in Fig.\ref{fig:fig1}b, where each atom is resolved with $>99.5\%$ fidelity, providing pristine access to the system's counting statistics. We acquire on the order of 700 to 1400 images, depending on the preparation, and extract the probability $P_N$ of finding $N$ atoms within a circular region of radius $R$. 

In Fig.\ref{fig:fig1}c, we show the measured probabilities $P_N(k_{\rm F} R)$ for $N = 0$ to $5$ for a representative sample with intermediate interaction strength ($\eta = 3.7(2)$) and reduced temperature $T/T_{\rm F} = 0.15(1)$. We compare our measurements to theoretical predictions from the Fredholm determinant formalism. The kernel associated to 2D free fermions is equal to the field correlation function $K(\bm x,\bm x')=\langle\hat{\Psi}^\dagger(\bm x)\hat{\Psi}(\bm x')\rangle$ with $\hat{\Psi}^\dagger(\bm x)$ and $\hat{\Psi}(\bm x)$ the fermionic field operators. We write $K(\bm{ x}, \bm{ x}')= k_{\rm F}^2 \tilde K(\bm \rho=k_{\rm F} \bm{ x}, \bm \rho'= k_{\rm F} \bm{x}')$, and leverage rotational symmetry to obtain an angular decomposition of $\tilde{K}$
using polar coordinates $\bm{\rho} \equiv (\rho,\phi)$ \cite{gouraud2022}:
\be \label{decomp}
\tilde K({\boldsymbol \rho}, {\bm \rho'}) =  \frac{1}{2 \pi \sqrt{\rho \rho'}} \sum_{\ell \in \mathbb{Z}} 
e^{ i \ell (\phi-\phi')} K_\ell(\rho,\rho') \;,
\ee 
where the kernels $K_{\ell}$ are given by
\begin{equation} \label{def_kernel_ell}
K_{\ell}( \rho,\rho') = \sqrt{\rho \rho'} \int_0^{+\infty} {\rm d} v \,  \frac{v\, J_\ell(v \rho)J_\ell(v \rho')}{(e^{1/t}-1)^{-1} e^{v^2/t} + 1}\;,
\end{equation}
with $t=T/T_{\rm F}$ and $J_\ell(x)$ the Bessel function of index $\ell$.
Using this decomposition we have obtained the following exact formula:
\begin{eqnarray} 
&&    P_N(r) = \frac{1}{2\pi}\int_0^{2\pi} {\rm d} \theta \, e^{-i N \theta} \prod_{\ell\geq 0}  F_\ell(\theta,r)^{\gamma_2(\ell)} 
    \quad , \quad \label{Pnintro} \\
&&    F_\ell(\theta,r) =
    {\rm Det} \left(1 - (1- e^{i \theta}) {\cal P}_r K_{\ell}\right) \;, \label{Detintro}
\end{eqnarray} 
 where $r=k_{\rm F} R$.
In Eq.\,(\ref{Detintro}) the notation ${\rm Det}$ denotes a Fredholm determinant,
${\cal P}_{r}(x,y)$ denotes the projector on the interval $x,y \in [0,r]$, $\gamma_2(0)= 1$ and $\gamma_2(\ell \geq 1) = 2$ (see Methods for more details). This Fredholm determinant depends on temperature through the Fermi factor in Eq.\,(\ref{def_kernel_ell}). At zero temperature, this determinant coincides with the one describing  the fluctuations of the smallest eigenvalues of large random matrices belonging to the Wishart-Laguerre ensemble of RMT, indexed by the angular momentum $\ell$ \cite{forrester2010}. In the formula for $P_N(r)$ (Eq.~(\ref{Pnintro})), the product over $\ell$ arises from the independence of the fluctuations in each $\ell$-sector. At very high temperature $T/T_{\rm F} \gg 1$, our result crosses over to the expected distribution for a Poisson point process
${P}^{\rm Poisson}_N(r) = \frac{r^{2 N}}{2^{2 N} N!} e^{- r^2/4}$.

 \begin{figure}[!t]
    \centering
    \includegraphics[width=\columnwidth]{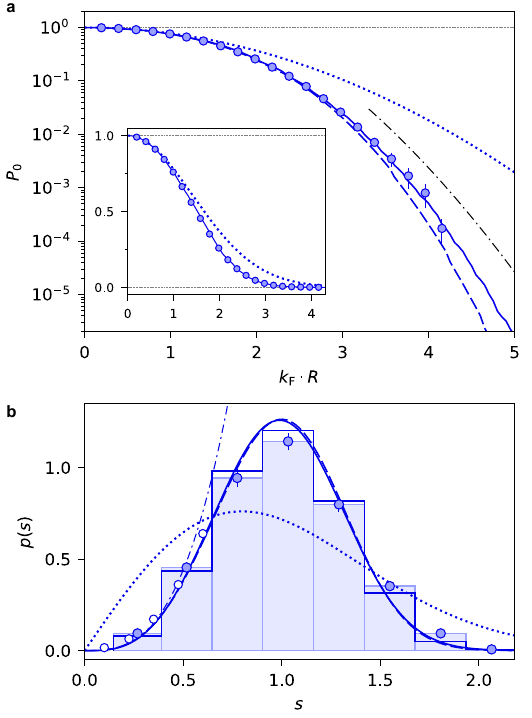}
    \caption{\textbf{Hole probability and nearest neighbor spacing distribution in a 2D Fermi gas}. Both quantities are extracted from samples of reduced temperature $T/T_{\rm F} = 0.15(1)$. Data is compared to theoretical predictions for $T= 0$ (blue dashed line) and $T/T_{\rm F} = 0.15$ (solid blue line), and the high-temperature Poisson behavior (dotted blue line). \textbf{(a)} Experimentally measured hole probability $P_0$ as function of $k_{\rm F}\cdot R$ (circles), shown in semi-logarithmic (main panel) and linear (inset) scale. Error bars represent the standard deviation of measurements taken across different probe disk locations. Also shown is the 1D asymptotic behavior $P_0 (r\rightarrow \infty) \propto e^{-r^2/2}$ (black dash-dotted line). \textbf{(b)}~The measured nearest-neighbour spacing distribution is represented as a light-blue histogram and blue circles, and compared to the predicted histogram for the same binning. The short range measurement extracted from the density-density correlation function Eq.~\eqref{approx_nns}  is shown in white circles, and the dash-dotted line shows a cubic fit to the short-range behavior of the theoretical prediction at $T=0$.}
    \label{fig:fig2}
\end{figure}

We find excellent agreement between our measurements and the above predictions with no fitting parameters, using the independently determined temperature $T/T_{\rm F} = 0.15(1)$ as sole input. A similar degree of agreement is observed for other interaction strengths considered in this work, including $\eta = 7.8(5)$ and $\eta = 2.1(2)$, though the latter exhibits the onset of small but measurable deviations (see Extended Data). These results reveal an unexpected breadth of applicability of RMT, since its  predictions can accurately describe spatial statistics of a Fermi gas well in the interacting regime.

In Fig.~\ref{fig:fig2}a, we focus on the hole probability $P_0$, which quantifies the likelihood of finding no particles within a given region of radius $R$. This is a key quantity for the description of spatial statistics, which characterize the rigidity of the Fermi gas under Pauli exclusion, with connections to the spacing distribution. The comparison of the experimentally measured $P_0(r)$ to finite-temperature predictions obtained from Fredholm determinant shows excellent agreement over several orders of magnitude in probability, down to values as small as $10^{-4}$. The precision of our experimental data in the tail of the distribution allows us to resolve the minute deviations from the predicted asymptotic form at $T=0$, $P_0(r)\sim e^{-r^3/9}$, valid for large radii~\cite{gouraud2022}.

Figure~\ref{fig:fig2}b shows the distribution of the nearest-neighbor spacing (NNS) $s$, a central quantity in the theory of random point processes. The spacing $s$ denotes the nearest-neighbor distance rescaled by its average $\bar{r}$, such that $\bar{s}=1$ by convention. The NNS distribution lies at the heart of Wigner and Dyson's seminal work on eigenvalue statistics in random matrices \cite{wigner1951,dyson1962,dyson1962a,mehta1991,forrester2010}. In one dimension, an ensemble of non-interacting fermions can be mapped onto the Gaussian Unitary Ensemble (GUE) of random matrices \cite{eisler2013,marino2014,dean2016}, where the level repulsion  corresponds to the spatial rigidity of the Fermi gas, with a spacing distribution well approximated by the Wigner surmise $p(s) \sim s^2 e^{-\kappa s^2}$ \cite{mehta1991}.

\begin{figure*}[!t]
    \centering
    \includegraphics[width=\textwidth]{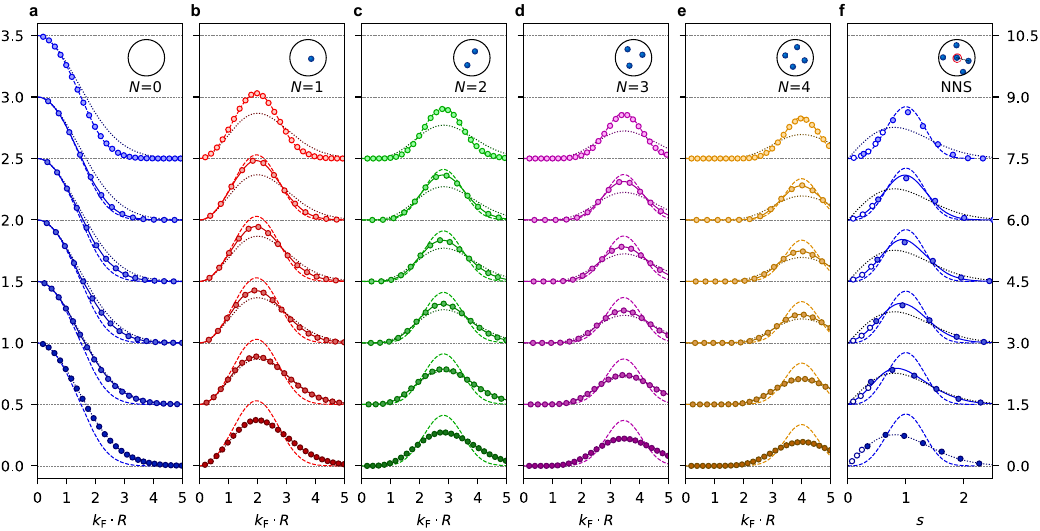}
    \caption{\textbf{Temperature crossover of the Counting Statistics.} \textbf{(a-f)} From top to bottom: data corresponding to samples of reduced temperatures $T/T_{\rm F}= [0.15(1),0.30(2), 0.53(3), 0.77(5), 2.1(2), \sim 20]$ and vertical ground state populations $p_0=[97.4(9)\%, 92.3(2)\%, 87.2(3)\%, 83.3(4)\%, 72.5(1.2)\%, \sim 23\%]$. Shown are the probability to find 0 \textbf{(a)}, 1 \textbf{(b)}, 2 \textbf{(c)}, 3 \textbf{(d)}, and 4 \textbf{(e)} atoms in a circular probe volume of variable radius $k_{\rm F}R$, as well as the nearest-neighbor spacing distribution (blue disks) as function of the scaled spacing $s$ \textbf{(f)}, including the short-range behavior (white disks) obtained from the measured density-density correlation function Eq.~\eqref{approx_nns} (see text). Like in figure \ref{fig:fig2}{\bf b}, the measured NNS distributions are obtained from histograms of measured spacings, but we only show the mean probability of each bin for legibility. Dashed lines show the zero temperature predictions and dotted lines the high-temperature Poissonian behavior. In the $2^{\rm nd}$ to $5^{\rm th}$ rows, solid lines present finite temperature numerical results, calculated independently using the experimentally determined reduced temperatures $T/T_{\rm F}$ and vertical state populations $p_\nu$. The data in the first row corresponds to samples with $\eta=3.7(2)$, and were already displayed in Figs.\,\ref{fig:fig1} and \ref{fig:fig2}.} 
    \label{fig:fig3}
\end{figure*}

In two dimensions, where RMT predictions are less established \cite{akemann2015,lacroix-a-chez-toine2019,akemann2022}, important differences emerge \cite{torquato2008}. First, the condition $p(s=0)=0$ does not imply level repulsion, but needs to be replaced by the stronger condition $p(s \rightarrow 0^+) \sim s^\alpha$ with $\alpha > 1$. It is indeed straightforward to show that for uncorrelated 2D particles (with Poisson number distribution), $p(s) \sim s e^{-\kappa s^2}$. Second, the tail of the NNS distribution in 2D does not follow a Gaussian decay, but a faster, non-Gaussian form directly related to the asymptotics of the hole probability. The experimentally measured NNS clearly deviates from the Poisson prediction, exhibiting strong short-range repulsion consistent with $p(s) \sim s^3$, which is the expected low temperature scaling $s^{d+1}$ with $d$ the dimension of the system \cite{torquato2008}. For relatively small spacings ($s\ll1$) we have also derived the approximate result:
\be \label{approx_nns}
p(s)\simeq \frac{\bar{r}^2}{2}sg_2(s\bar{r}),
\ee
where $g_2$ is the reduced density-density correlation function (see Methods). Using this relation and the measured $g_2$ (see Extended Data), we obtain an independent measurement of the NNS at short distance, yielding additional data points (see Fig.~\ref{fig:fig2}b). The NNS distribution also displays the expected faster decay at large $s$, compared to the 1D case. We compare our data to the NNS distribution computed from the finite-temperature Fredholm determinant calculations,
which is given by  
$p(s)=- \bar r \frac{ d P^{}_{\odot}}{dr}|_{r=s \bar r}$,
where $P^{}_{\odot}(r)$  is the probability of finding no atom in a disk of radius $r$ conditioned on one atom being at its center, for which we have derived an exact expression (see Methods). For accurate comparison, we also compute the histogram derived from the theoretical result using the same binning as the data. We find very good agreement without any free parameters. This constitutes, to the best of our knowledge, the first direct measurement of the spacing statistics in a 2D quantum gas. In ultracold systems, the spacing statistics of a 1D point process was observed in the collisional spectra of Erbium atoms \cite{frisch2014}. 

We now turn to the study of the temperature dependence of the counting statistics of our system. The hole probability and the NNS characterize the rigidity of the Fermi gas. 
Our ability to prepare samples of variable degeneracy~\cite{dixmerias2025} allows us to witness the breakdown of this rigidity as these observables transition to a classical behavior.
In order to disentangle interaction and temperature effects, we consider samples prepared with vanishingly weak attraction with $\eta=20.5(10)$. All samples are initially evaporatively cooled in the optical trap, and contain $\sim 150$ atoms per spin state. After a thermalization time of 1.6\,s, the trap depth is adiabatically ramped up to $\sim650$\,nK, corresponding to a vertical trapping frequency $\omega_z = 2\pi \times 3.0(1)$\,kHz. We then modulate the trap intensity with a frequency $\simeq 2\,\omega_z$ and a relative amplitude of 5\,\% for a variable time. Following this controlled heating, the atoms are left to thermalize for 1.8\,s, before one of the two spin components is expelled. The resulting single-component systems realize (quasi-) two-dimensional non-interacting Fermi gases, with fixed trapping parameters and similar atom numbers, and a final temperature set by the modulation time. We additionally prepare a classical (hot) gas to serve as a reference (see Ref. \cite{dixmerias2025} for more details). 

In the vertical direction the atomic motion is quantized, with a quantum of vibration corresponding to a temperature scale $T_z=\hbar\omega_z/k_{\rm B} =143(3)$\,nK. For the previously discussed (interacting) samples, $T$, $T_{\rm F}$, the interaction energy, and $T_z$ were chosen to ensure an occupation of the excited $z$-level states below 4\%, such that the samples could be considered essentially 2D. As we now explore a large dynamic range in temperature, we consequently expect the fraction $p_\nu$ in the $\nu$-th excited $z$--level to increase with temperature. The system can then be described as a superposition of independent 2D systems with Fermi temperature $p_\nu T_F$ and Fermi wavevector $\sqrt{p_\nu}k_F$. In this case the theoretical prediction for the hole probability for quasi-2D free fermions
becomes (see Methods):
\be 
{P}^{\rm q2D}_0(r, t)   = \prod_{\nu \geq 0} {P}_0 \left( r \sqrt{p_\nu}, t/p_\nu\right),
\ee 
with ${P}_0(r,t)$ the 2D result given in Eq. \eqref{Pnintro}. 
We obtain similar formulas for ${P}^{\rm q2D}_N(r, t)$ for $N \geq 1$.

In Fig.~\ref{fig:fig3}, we report the measurement of the probabilities to find $N=0, ...,4$ in a circle of radius $R$, and the next-nearest-neighbor $p(s)$, for a selection of samples. For each reported quantity we find again excellent quantitative agreement with the predictions based on Fredholm determinant without any fitting parameters. These results provide an accurate measurement of the smooth crossover from the correlated regime driven by Pauli's exclusion at low temperature to a Poissonian regime at high temperature.

We have experimentally investigated number and spacing statistics in 2D degenerate Fermi gases with continuum quantum gas microscopy. The near perfect detection allowed us to finely measure the behavior of the hole probability and nearest-neighbor spacing in an interacting Fermi gas near zero temperature. The measurements agree remarkably well with exact computations based on Fredholm determinants, using the connection between non-interacting 2D fermions and the Wishart-Laguerre ensemble of Random Matrix Theory. We further characterized the crossover in the number and spacing statistics from the low-temperature distributions exhibiting Pauli repulsion to the classical, Poissonian behavior with increasing temperature.
Our results constitute to our knowledge the first experimental observation of a Fermi-sphere point process. Notably, our data agrees well with RMT-based computations even in regimes of relatively strong interactions. We believe that this unexpected result may be explained by the similar resilience to interactions of same-spin density correlation functions in interacting Fermi gases discovered in a recent work~\cite{daix2025}. Indeed, the same correlation functions appear in the expansion of the characteristic function from which the full counting statistics are derived. While this result remains to be established analytically, our work underscores the importance of counting statistics observables and the tools of RMT for the study of strongly correlated systems.\\

\textbf{Acknowledgements:} We thank Gabriel Gouraud for stimulating discussions. T.Y. is grateful to Antoine Heidmann for his crucial support as head of Laboratoire Kastler Brossel. This work has been supported by Agence Nationale de la Recherche (Grant No. ANR-21-CE30-0021 and No. ANR-23-CE30-0020-01), CNRS (Tremplin@INP 2020), and R{\'e}gion Ile-de-France in the framework of DIM SIRTEQ and DIM QuanTiP.\\

\textbf{Author contributions:}  M.D., C.D., J.V., and T.d.J. performed the experiment. M.D., C.D., J.V., T.d.J., and B.P., all contributed to the analysis of the data. G.d.V.d.V., P.L., and G.S. contributed to the theoretical analysis and carried out the Fredholm determinant calculations. T.Y. planned and supervised the study. All authors contributed to the interpretation of the results and to the writing of the manuscript.\\

\textbf{Competing interests:} The authors declare no competing interests.

\section*{Methods} 

\noindent \textbf{Continuum Quantum Gas Microscopy.} We probe the system by freezing the motion of the atoms initially evolving in continuous space, before
imaging their positions. The pinning process consists in turning on a deep optical lattice within $\sim 10\,\mu$s, which traps each atom in the nearest potential well \cite{verstraten2025,dejongh2025}. The ramp duration is carefully chosen: it is slow enough to avoid projecting atoms into high lattice bands, yet fast enough to prevent significant motion before pinning is complete. The pinning lattice is generated by the self-interference of a red-detuned 1064\,nm laser, with three beams crossing at $120^\circ$ angles in the horizontal plane to form a triangular lattice with a spacing $a_{\rm L} = 709\,$nm  \cite{jin2024}. The lattice wells have a measured trapping frequency of $\sim1$\,MHz. Immediately after pinning, we apply Raman sideband cooling \cite{verstraten2025}. This serves a dual purpose: it cools the atoms close to the vibrational ground state of their respective lattice wells and simultaneously induces fluorescence. The spontaneously emitted photons are collected by a high-resolution objective and imaged onto a CCD camera. The resulting fluorescence images are processed using a highly accurate neural network algorithm to pinpoint the location of each atom \cite{verstraten2025}.\\

\noindent \textbf{Counting statistics measurement.} We compute the counting statistics from the positions of the detected atoms in each image. A random point in space is selected, around which disks of variable radius $R$ are drawn. Within each disk, we count the number of atoms across all images. From this we obtain the particle number counting statistics as function of $k_{\rm F}R$, where $k_{\rm F}$ is determined from the average density. Confidence intervals are determined by repeating the measurement across 100 randomly drawn disk center positions. We restrict our analysis to a quasi-homogeneous region at the cloud center of the cloud, where density variations are limited to $\pm 5\,\%$.\\

\noindent \textbf{NNS measurement.} To compute the nearest neighbor distance (NNS) distribution, we loop over the atomic positions in the central region and store distance that separates each considered atom from its nearest neighbor. We repeat the measure for all the images of a given preparation. From this we extract an histogram that, after normalization, yields the NNS probability distribution. Errorbars are obtained via bootstrapping. The short distance data points obtained via Eq.~\eqref{approx_NNS}, uses a spline to the experimentally measured $g_2$-functions shown in the Extended Data. The final result is discretized for readability.\\

\noindent {\bf 2D Model}. The model consists of non-interacting fermions in 2D with single particle Hamiltonian 
$H_0={\bf p}^2/(2 m)$ where ${\bf p}$ is the momentum. We consider here the grand canonical ensemble at temperature 
$T$ and
chemical potential $\mu$. One defines by convention $k_{\rm F} = \sqrt{4 \pi n}$ and $T_{\rm F} =2 \pi n \hbar^2/(m k_B)$,
where $n$ is the mean density (this corresponds to the standard definition only at $T=0$), These
parameters are related through $e^{\mu/(k_B T)} = e^{1/t}-1$ and $t=T/T_F$.
\\

\noindent {\bf Quasi-2D Model.} The model consists of non-interacting fermions in 3D strongly
confined in the transversal direction ($z$) by a harmonic potential, 
with single particle Hamiltonian 
$H_0={\bf p}^2/(2 m) + \frac{1}{2} m \omega_z^2 z^2$. One defines by convention
$k_{\rm F} = \sqrt{ 4 \pi n }$ and $T_{\rm F} = \hbar^2 k_{\rm F}^2/(2 m k_B)$.
Here $n$ denotes the total 2D density (i.e. the 3D density integrated over $z$),
which is given by
\be 
 n = \frac{k_B T m}{2 \pi} \sum_{\nu\geq 0} \log( 1 +  e^{(\mu - \hbar \omega_z \nu)/(k_B T)} ) \;.
\ee 
One defines the occupation fraction of level $\nu$
\be 
p_\nu = \frac{ \log( 1 +  e^{(\mu - \hbar \omega_z \nu)/(k_B T)} )  }{ \sum_{\nu \geq 0} 
\log( 1 +  e^{ (\mu - \hbar \omega_z \nu)/(k_B T)} )  }
\ee 
In this model we will redefine $r = k_{\rm F} R$ and $t= T/T_F$. 
\\

\noindent {\bf Counting formalism in 2D.}
We call ${\cal N}_R$ the number of fermions inside a given disk of radius $R=r/k_{\rm F}$
One defines the probabilities 
${P}_N(r)$ that there are $N$ fermions in the disk, i.e. that ${\cal N}_R=N$,
as well as the generating function (GF), i.e. the full counting statistics (FCS), 
\be  \label{GF_NR}
G_R(\sigma) = \langle e^{-\sigma {\cal N}_R} \rangle = \sum_{N \ge 0} {P}_N(r) e^{-\sigma N} \;
\ee
where $\langle \cdots \rangle$ denotes the quantum grand-canonical average. Once the GF is known one can retrieve the probabilities using Cauchy's theorem from complex analysis as
\be 
P_N(r) = \frac{1}{2\pi} \int_0^{2\pi} {\rm d} \theta \, e^{-i \theta N} G_R(-i\theta) \;.
\ee 
Since the positions of non-interacting fermions in the grand canonical ensemble 
form a determinantal point process, we obtain the following explicit expression
for the generating function
\be \label{det_GF}
G_R(\sigma) = {\rm Det}(1 - (1-e^{-\sigma})\Pi_R K )
\ee 
in terms of the 2D kernel $K({\bf x}, {\bf x'})$ defined in the main text and where $\Pi_R({\bf x},{\bf y})$ denotes the projector on the disk centered at the origin and of radius $R$. 
Using the angular decomposition and extending the methods of \cite{gouraud2022},
to finite temperature the generating function can be expressed as
the product of Fredholm determinants 
which appears in the right hand side of \eqref{Pnintro} in the main text.
\\

\noindent {\bf Counting formalism in quasi-2D.} In the quasi-2D model the
total number of fermions in the cylinder of radius $R=r/k_F$ is denoted 
${\cal N}_R$, and the corresponding probabilities that ${\cal N}_R=N$
are denoted $P^{\rm q2D}_N(r,t)$, to distinguish them from their 2D analog $P_N(r,t)$. 
The random variable ${\cal N}_R$ 
can be written as the sum ${\cal N}_R = \sum_{\nu \geq 0} {\cal N}_{R,\nu}$
of independent random variables ${\cal N}_{R,\nu}$ for an effective 2D gas
where $\mu$ has been shifted as $\mu \to \mu - \nu \hbar \omega_z$. 
Let us first introduce the hole probability corresponding to the $\nu$-th $z$-level, which reads:
\be \label{hole_proba_q2d}
P_0^{\nu}(r,t)= P_0\left( r \sqrt{p_\nu}, \frac{T}{T_{\rm F} p_\nu} \right) 
\ee taking into account the rescaled density $p_\nu n$, with associated Fermi momentum and temperature given respectively by
$\sqrt{p_\nu}k_{\rm F}$ and $T_{\rm F} p_\nu$. Since the various $z$-levels are independent, the hole probability for the whole cylinder is simply the product $P^{\rm q2D}_0(r,t) = \prod_{\nu \geq 0} P_0^{\nu}(r,t)$. A similar formula holds for the quasi-2D generating function 
\be \label{fcs} 
\sum_{N\geq 0} e^{-\sigma N} {P}^{\rm q2D}_N(r,t) = \prod_{\nu \geq 0} \left(  \sum_{N \geq 0 } e^{- \sigma N } 
P_N^\nu \left( r, t \right)  \right) 
\ee  
where $P_N^\nu(r,t)$ denotes the probability to find $N$ atoms in the level $\nu$ (similarly to the hole probability in Eq.\,\eqref{hole_proba_q2d}). This expression allows to extract the quasi-2D probabilities ${P}^{\rm q2D}_N(r,t)$.
\\

\noindent{\bf NNS formula in 2D.} 
Let us introduce $P_\odot(r)$, the probability that, around a point of the point process chosen at random, 
there are no other point within a distance $R=r/k_F$. The PDF of the
nearest neighbor spacing (NNS) defined in the text can be expressed as $p(s)= - \bar r \frac{d}{d r} P_\odot(r)\vert_{r = s \bar{r}}$
where $k_{\rm F} \bar r$ is the mean spacing. The probability $P_\odot(r)$
can be constructed from a hole probability
as follows. 
We denote 
$\tilde{P}_0(r,\epsilon)$ the probability there are no
fermions inside an annulus $\epsilon < \rho < r/k_F$, where $\tilde{P}_0(r,0)=P_0(r)$. 
Then one can show \cite{scardicchio2009} that 
\begin{equation}\label{eq:nnprob_torq}
    P_\odot(r) = \lim_{\epsilon \to 0}\frac{\tilde{P}_0(r,\epsilon) - \tilde{P}_0(r,0)}{n \pi \epsilon^2} \;,
\end{equation} where $n \pi \epsilon^2$ for small $\epsilon$ coincides with the probability
that there is a fermion in a disk of radius $\epsilon$ around the origin.
Since for 2D non-interacting fermions $\tilde{P}_0(r,\epsilon)$ is given by the 
same Fredholm determinant formula as in \eqref{det_GF} setting $s=+\infty$, 
where now $\Pi_R$ is replaced by the 
projector onto the annulus $\epsilon<\rho<R=r/k_F$, taking the limit $\epsilon \to 0$ using
standard identities for derivatives of Fredholm determinants, we obtain the exact formula:

\begin{equation}\label{eq:EP_fredholm}
    P_\odot(r) =4\pi P_0(r) \left[(1 - {\cal P}_r \tilde K {\cal P}_r)^{-1} {\cal P}_r \tilde K {\cal P}_r \right]({\bm 0},{\bm 0})\,
\end{equation} where the operator on the r.h.s. should be evaluated at $(\bm{\rho},\bm{\rho}')=({\bm 0},{\bm 0})$, and
\be  \label{def_hat_K}
\tilde K( \bm{\rho}, \bm{ \rho} \,')
= \frac{1}{2 \pi} \int_0^{+\infty} {\rm d}v \frac{v \, J_0(v |\bm{ \rho} - \bm{ \rho} \,'|)} {(e^{1/t}-1)^{-1} e^{v^2/t} + 1} \;.
\ee

A small spacing approximate formula for the NNS may be derived from two equivalent exact definitions~\cite{torquato2008}:
\bea
\tilde p(r) &=& - \frac{d}{d r} P_\odot(r) \\
&=& \frac{r}{2} G^{(c)}_2(r)\, P_\odot(r), \label{approx_NNS}
\eea
where $G^{(c)}_2(r)dr$ is the conditional probability of finding an atom in the circular shell centered on one atom of radius $r$ and width $dr$. In the short-range regime $r\ll 1$, the probability of having more than one atom in the circular shell is small, and we have $G^{(c)}_2(r)\simeq g_2(r)$, the usual two-point reduced density correlation function. Approximating further $P_\odot(r)\simeq 1$ for small $r$ leads to
\bea
\tilde p(r)\simeq\frac{r}{2}g_2(r)
  \label{approx_NNS_final}
\eea
or equivalently Eq.\,\eqref{approx_nns} in the main text. This leads to the expansion:
\bea
\tilde p(r)= \frac{r^3}{8}+\frac{5 r^5}{384} + O(r^7)\,.
\eea At $T=0$, the discrepancy between the exact result for $\tilde p(r)$ and the approximation is only $O(r^7)$. 
\\

\noindent{\bf NNS formula in quasi-2D.} For the quasi-2D model the formula (\ref{eq:nnprob_torq}) still applies 
for the corresponding probability $P_\odot^{\rm q2D}(r)$, 
where now $\tilde{P}_0(r,\epsilon)$ denotes the hole probability associated to the cylindrical shell $\{\epsilon < \rho < r\, ,\, z\in \mathbb{R}\}$.
Using, as above, that such a hole probability, is a product over hole probabilities associated to each level $\nu$ 
one obtains in the limit $\epsilon \to 0$
\be 
P_\odot^{\rm q2D}(r,t)={P}^{\rm q2D}_0(r,t)  \sum_{\nu \geq 0} p_\nu\frac{ P_\odot(r \sqrt{p_\nu},t p_\nu) }{ P_0( r \sqrt{p_{\nu}},t p_{\nu}) } \,,
\ee yielding the NNS $p^{\rm q2D}(s)= - \bar r \frac{d}{d r} P_\odot^{\rm q2D}(r)\vert_{r = s \bar{r}}$.
\\

\noindent \textbf{Numerical evaluations.} The Fredholm determinants (FD) in Eqs. (\ref{Pnintro}) and (\ref{Detintro}) have been computed numerically 
as functions of $r,t,N$, 
using Bornemann's method~\cite{bornemann2010,bornemann2011} where the kernel is approximated by
a standard matrix with proper weights. The main idea of the algorithm 
goes as follows: we truncate the infinite product over $\ell$ in (\ref{Pnintro}) at $\ell_{\max}$ and, for $\ell\in[0,\ell_{\rm max}]$, we compute the FD $F_\ell(\theta,r)$ on a truncated 
interval $[0,L_2]$ with $r \ll L_2$. 
For this purpose, 
we pre-compute the kernel matrices $\hat K^G_\ell$ of sizes $M_2$, with
matrix elements
\begin{equation}\label{eq:numerical_kernel}
    (\hat K_\ell^{\rm G})_{i,j} \approx \sqrt{\gamma_i x_i \gamma_j x_j }  \sum_{s=1}^{M_1} \eta_s v_s 
 \frac{J_\ell(v_s x_i) J_\ell(v_s x_j )}{1 + (e^{1/t}-1)^{-1} \;, e^{v_s^2/t }} 
\end{equation}
where $(x_i,\gamma_i)$, $i=1,\dots ,M_2$ are the nodes and the weights of a Gaussian quadrature rule (GQR) for $x \in [0,L_2]$.
We have introduced an upper cutoff $L_1$ to evaluate numerically
the integral over $v$ in \eqref{def_kernel_ell}. In (\ref{eq:numerical_kernel}) the $(v_i,\eta_i)$, for $i=1,\dots ,M_1$, denote the nodes and the weights of a GQR for $v \in [0,L_1]$. 

From these matrices $\hat K^{\rm G}_\ell$, we compute for fixed $r$ and $\theta$, the standard determinant
of size $M_2$
\begin{equation}\label{eq:numerical_F_hat}
        F_\ell(\theta,r)\approx \det\left( 1 + (1- e^{i\theta}){\cal P}_{r}\hat K^{\rm G}_\ell\right) \;.
\end{equation}
Here ${\cal P}_r$ is the diagonal matrix with elements $({\cal P}_r)_{i,i} = \Theta(x_i) \Theta(r-x_i)$ with $\Theta(x) = 1$ if $x\geq 0$ and $\Theta(x)=0$ otherwise. At the end we perform the product over $\ell$ and the numerical integration over $\theta$ 
using Gaussian quadrature rule (GQR) for each $N=0,1,\dots$ to obtain a numerical evaluation of $P_N(r)$ in (\ref{Pnintro})-(\ref{Detintro}). The numerical evaluation of \eqref{eq:EP_fredholm} proceeds similarly, performing a matrix inversion involving the discrete kernel \eqref{eq:numerical_kernel}. Performing a numerical derivative gives the NNS distribution plotted in Fig. \ref{fig:fig2}b. The calculations for the quasi-2D model use the $p_\nu$ as input (table I) 
are similar using the formula \eqref{fcs}, taking only into account the level with non-negligible population. 

In our simulations we have chosen the following parameters for which we obtain excellent convergence: $\ell_{\rm max}=10$, $M_1=2500$, $L_1=1000$, $M_2=1000$, $L_2 = 20$, $M_3 = 2500$. To obtain the data showed in the plots we have discretized the variable $r$ as $r_i = i \Delta r$, with $\Delta r = 0.05$ and $r_{\rm max}=8$ and we have run the above procedure on a separate core on the cluster for each $r$ independently.

\newpage

\renewcommand\thefigure{S\arabic{figure}}
\renewcommand\theHfigure{S\arabic{figure}}
\setcounter{figure}{0} 

\renewcommand\theequation{S\arabic{equation}}
\renewcommand\theHequation{S\arabic{equation}}
\setcounter{equation}{0}

\begin{widetext}
\newpage
\pagebreak

\section*{Extended Data}

\newcolumntype{s}{>{\columncolor[HTML]{DDDDDD}} c}
\begin{table*}[!h]
\begin{center}
\begin{tabular}{ |c|s|c|s|c|c|c|c|c| }
\hline
$\eta$       & 2.1(2)   &  3.7(2) &  7.8(5) & 20.8(6) & 20.7(6) &  20.6(6) &  20.4(6)  & 20.2(7)
\\ \hline\hline
$T/T_{\rm F}$& 0.11(1)  & 0.15(1) & 0.17(3) & 0.30(2) & 0.53(3) & 0.77(5) & 2.06(14) & $\sim 20$
\\ \hline
$T/T_{z}$    & 0.07(1)  & 0.09(1) & 0.10(1) & 0.23(1) & 0.36(1) & 0.45(1) & 0.72(2)  & $\sim 4$
\\ \hline \hline
$p_0\,(\%)$  & 96.9(1.2)& 97.4(9) & 96.8(9) & 92.2(8)& 87.2(8)& 83.3(9)& 72.6(1.4)& $\sim 23$
\\ \hline
$p_1\,(\%)$  &3.1(1.2)  & 2.6(9)  & 3.2(9)  & 7.7(8)  & 12.0(7) & 14.7(6) & 20.4(7)  & $\sim 18$
\\ \hline
$p_2\,(\%)$  & $< 0.1$  & $< 0.1$ & $< 0.1$  & $< 0.1$ & 0.8(1)  & 1.7(2)  & 5.3(5)   & $\sim 14$
\\ \hline
$p_3\,(\%)$  & $< 0.1$  & $< 0.1$ & $< 0.1$  & $< 0.1$ & $< 0.1$ & 0.2(1)  & 1.3(2)  & $\sim 11$
\\ \hline
\end{tabular}
\caption{Relevant quantities of the preparations presented in Fig.\,\ref{fig:fig3} of the main text (uncolored columns) and in Extended Data Fig.\,\ref{fig:ex_fig1} and Fig.\,\ref{fig:ex_fig2} (grey columns).}
\label{tab:fig3_quantities}
\end{center}
\end{table*}

\newpage

\begin{figure*}[ht!]
\centering
\includegraphics[width=0.8\textwidth]{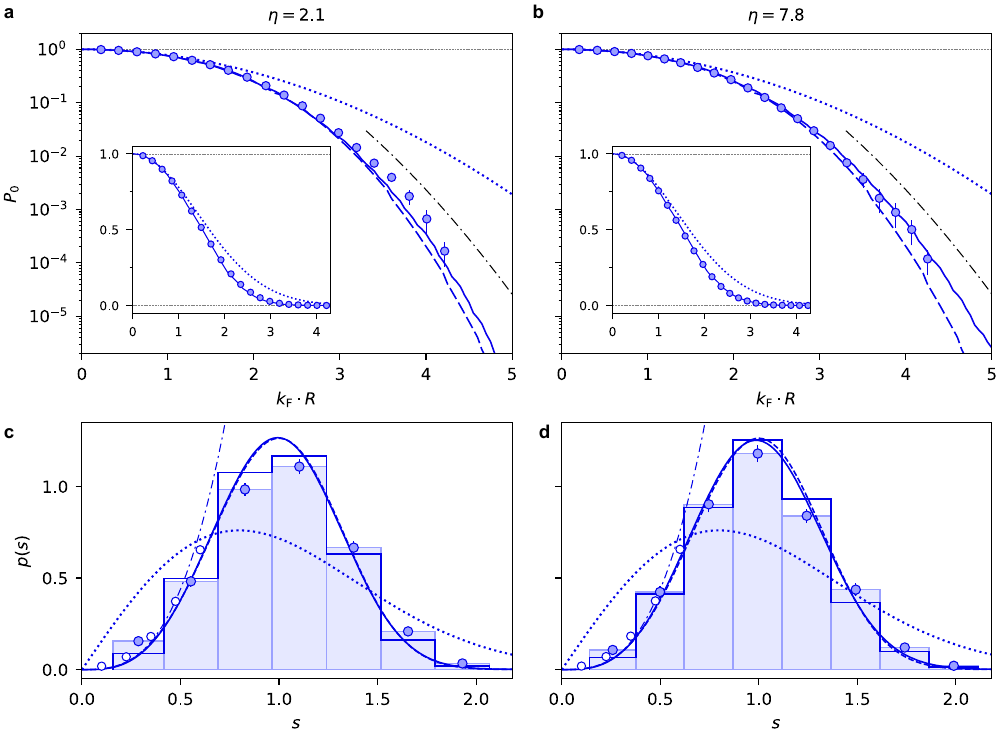}
\caption{\textbf{Additional hole probabilities and nearest neighbor spacing distributions for 2D Fermi gases}. (\textbf{a}-\textbf{b})  Experimental  measurement of the hole probability $P_0$ (blue markers) as a function  of $k_{\rm F}R$ for samples at interaction strength $\eta = 2.1(2)$ (\textbf{a}) and $\eta = 7.8(5)$ (\textbf{b}), respectively, with reduced temperatures $T/T_{\rm F}=0.11(1)$ and $T/T_{\rm F}=0.17(3)$, presented in semi-logarithmic (main  panels)  and  linear (insets) scales. Dashed (resp. solid) blue lines are theoretical predictions at zero temperature (resp. experimentally measured temperatures). Additional curves show the high-temperature Poisson behavior (dotted blue lines) and the zero temperature asymptotic behaviors in 1D (black dot-dashed lines). (\textbf{c}-\textbf{d}) Histograms (light-blue), combined with data points (blue circles), of the nearest-neighbour spacing distributions $p(s)$ for $\eta = 2.1(2)$ (\textbf{c}) and $\eta = 7.8(5)$ (\textbf{d}). The corresponding finite temperature predictions are shown with blue solid lines and histograms using similar binning. We present the classical Poisson behavior (blue dotted curves) and the zero temperature prediction for 2D fermions (blue dashed curve). The short range measurements extracted from the density-density correlation functions (Eq.\,\eqref{approx_nns}) is shown as white circles. The dash-dotted lines are a cubic fit of the zero temperature prediction at short range.}
\label{fig:ex_fig1}
\end{figure*}

\newpage

\begin{figure*}[ht!]
\centering
\includegraphics[width=\textwidth]{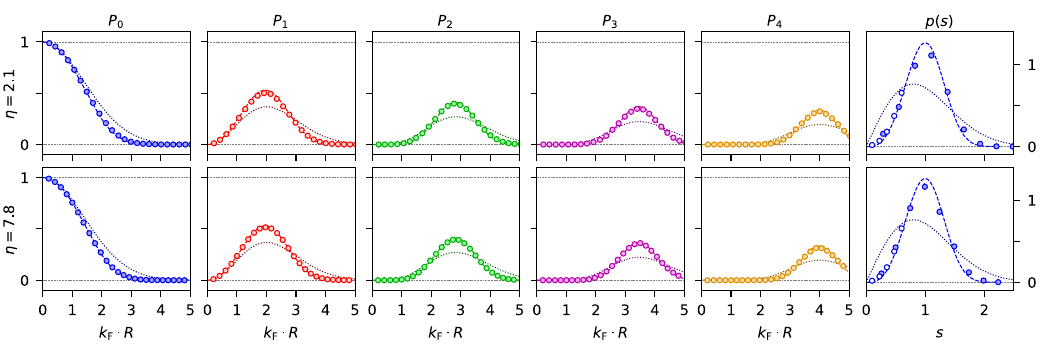}
\caption{\textbf{Additional counting statistics}. Data points are the probabilities $P_N$ to find $N=[0,1,2,3,4]$ atoms in a circular probe volume of variable radius $k_{\rm F}R$, as well as the nearest-neighbor spacing distribution $p(s)$ as function of the scaled spacing $s$ for $\eta = 2.1(2)$ (first row) and $\eta = 7.8(5)$ (second row), corresponding to $T/T_{\rm F}=0.11(1)$ and $T/T_{\rm F}=0.17(3)$ respectively. Dashed lines show the zero temperature predictions and dotted lines the high-temperature Poissonian behavior. The short-range behavior (white disks) obtained from the measured density-density correlation function Eq.\,\eqref{approx_nns} (see main text).}
\label{fig:ex_fig2}
\end{figure*}

\newpage

\begin{figure*}[ht!]
\centering
\includegraphics[width=0.8\textwidth]{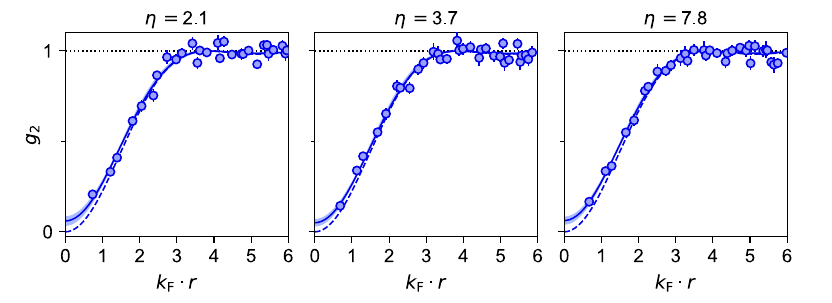}
\caption{\textbf{Two-point correlations}. The blue data points are experimental measurements of the same-spin density-density correlation function $g_2$ as a function of $k_{\rm F}r$ for $\eta = [2.1(2), 3.7(2), 7.8(5)]$ (see \cite{daix2025}). The blue dashed lines are the zero temperature prediction for the ideal Fermi gas. The blue solid lines are two-parameter fits (see \cite{daix2025}), which we use as splines to obtain the short-range behavior of the NNS distribution using Eq.\,\eqref{approx_nns} of the main text, with shaded areas indicating uncertainty.}
\label{fig:ex_fig3}
\end{figure*}
\end{widetext}


\begin{thebibliography}{40}%
\makeatletter
\providecommand \@ifxundefined [1]{%
 \@ifx{#1\undefined}
}%
\providecommand \@ifnum [1]{%
 \ifnum #1\expandafter \@firstoftwo
 \else \expandafter \@secondoftwo
 \fi
}%
\providecommand \@ifx [1]{%
 \ifx #1\expandafter \@firstoftwo
 \else \expandafter \@secondoftwo
 \fi
}%
\providecommand \natexlab [1]{#1}%
\providecommand \enquote  [1]{``#1''}%
\providecommand \bibnamefont  [1]{#1}%
\providecommand \bibfnamefont [1]{#1}%
\providecommand \citenamefont [1]{#1}%
\providecommand \href@noop [0]{\@secondoftwo}%
\providecommand \href [0]{\begingroup \@sanitize@url \@href}%
\providecommand \@href[1]{\@@startlink{#1}\@@href}%
\providecommand \@@href[1]{\endgroup#1\@@endlink}%
\providecommand \@sanitize@url [0]{\catcode `\\12\catcode `\$12\catcode
  `\&12\catcode `\#12\catcode `\^12\catcode `\_12\catcode `\%12\relax}%
\providecommand \@@startlink[1]{}%
\providecommand \@@endlink[0]{}%
\providecommand \url  [0]{\begingroup\@sanitize@url \@url }%
\providecommand \@url [1]{\endgroup\@href {#1}{\urlprefix }}%
\providecommand \urlprefix  [0]{URL }%
\providecommand \Eprint [0]{\href }%
\providecommand \doibase [0]{https://doi.org/}%
\providecommand \selectlanguage [0]{\@gobble}%
\providecommand \bibinfo  [0]{\@secondoftwo}%
\providecommand \bibfield  [0]{\@secondoftwo}%
\providecommand \translation [1]{[#1]}%
\providecommand \BibitemOpen [0]{}%
\providecommand \bibitemStop [0]{}%
\providecommand \bibitemNoStop [0]{.\EOS\space}%
\providecommand \EOS [0]{\spacefactor3000\relax}%
\providecommand \BibitemShut  [1]{\csname bibitem#1\endcsname}%
\let\auto@bib@innerbib\@empty
\bibitem [{\citenamefont {Wigner}(1951)}]{wigner1951}%
  \BibitemOpen
  \bibfield  {author} {\bibinfo {author} {\bibfnamefont {E.~P.}\ \bibnamefont
  {Wigner}},\ }\href {https://doi.org/10.1017/S0305004100027237} {\bibfield
  {journal} {\bibinfo  {journal} {Mathematical Proceedings of the Cambridge
  Philosophical Society}\ }\textbf {\bibinfo {volume} {47}},\ \bibinfo {pages}
  {790} (\bibinfo {year} {1951})}\BibitemShut {NoStop}%
\bibitem [{\citenamefont {Dyson}(1962{\natexlab{a}})}]{dyson1962}%
  \BibitemOpen
  \bibfield  {author} {\bibinfo {author} {\bibfnamefont {F.~J.}\ \bibnamefont
  {Dyson}},\ }\href {https://doi.org/10.1063/1.1703773} {\bibfield  {journal}
  {\bibinfo  {journal} {Journal of Mathematical Physics}\ }\textbf {\bibinfo
  {volume} {3}},\ \bibinfo {pages} {140} (\bibinfo {year}
  {1962}{\natexlab{a}})}\BibitemShut {NoStop}%
\bibitem [{\citenamefont {Dyson}(1962{\natexlab{b}})}]{dyson1962a}%
  \BibitemOpen
  \bibfield  {author} {\bibinfo {author} {\bibfnamefont {F.~J.}\ \bibnamefont
  {Dyson}},\ }\href {https://doi.org/10.1063/1.1703774} {\bibfield  {journal}
  {\bibinfo  {journal} {Journal of Mathematical Physics}\ }\textbf {\bibinfo
  {volume} {3}},\ \bibinfo {pages} {157} (\bibinfo {year}
  {1962}{\natexlab{b}})}\BibitemShut {NoStop}%
\bibitem [{\citenamefont {Mehta}(1991)}]{mehta1991}%
  \BibitemOpen
  \bibfield  {author} {\bibinfo {author} {\bibfnamefont {M.~L.}\ \bibnamefont
  {Mehta}},\ }\href@noop {} {\emph {\bibinfo {title} {Random Matrices}}},\
  \bibinfo {edition} {2nd}\ ed.\ (\bibinfo  {publisher} {Academic Press},\
  \bibinfo {address} {Boston},\ \bibinfo {year} {1991})\BibitemShut {NoStop}%
\bibitem [{\citenamefont {Forrester}(2010)}]{forrester2010}%
  \BibitemOpen
  \bibfield  {author} {\bibinfo {author} {\bibfnamefont {P.}~\bibnamefont
  {Forrester}},\ }\href {https://www.jstor.org/stable/j.ctt7t5vq} {\emph
  {\bibinfo {title} {Log-{{Gases}} and {{Random Matrices}} ({{LMS-34}})}}}\
  (\bibinfo  {publisher} {Princeton University Press},\ \bibinfo {year}
  {2010})\BibitemShut {NoStop}%
\bibitem [{\citenamefont {Akemann}\ \emph {et~al.}(2015)\citenamefont
  {Akemann}, \citenamefont {Baik}, \citenamefont {Francesco}, \citenamefont
  {Akemann}, \citenamefont {Baik},\ and\ \citenamefont
  {Francesco}}]{akemann2015}%
  \BibitemOpen
  \bibinfo {editor} {\bibfnamefont {G.}~\bibnamefont {Akemann}}, \bibinfo
  {editor} {\bibfnamefont {J.}~\bibnamefont {Baik}}, \bibinfo {editor}
  {\bibfnamefont {P.~D.}\ \bibnamefont {Francesco}}, \bibinfo {editor}
  {\bibfnamefont {G.}~\bibnamefont {Akemann}}, \bibinfo {editor} {\bibfnamefont
  {J.}~\bibnamefont {Baik}},\ and\ \bibinfo {editor} {\bibfnamefont {P.~D.}\
  \bibnamefont {Francesco}},\ eds.,\ \href@noop {} {\emph {\bibinfo {title}
  {The {{Oxford Handbook}} of {{Random Matrix Theory}}}}},\ Oxford
  {{Handbooks}}\ (\bibinfo  {publisher} {Oxford University Press},\ \bibinfo
  {address} {Oxford, New York},\ \bibinfo {year} {2015})\BibitemShut {NoStop}%
\bibitem [{\citenamefont {Castin}(2007)}]{castin2007}%
  \BibitemOpen
  \bibfield  {author} {\bibinfo {author} {\bibfnamefont {Y.}~\bibnamefont
  {Castin}},\ }in\ \href {https://doi.org/10.3254/978-1-58603-846-5-289} {\emph
  {\bibinfo {booktitle} {Lecture Notes of the 2006 {{Varenna Enrico Fermi
  School}} on {{Fermi}} Gases}}},\ \bibinfo {editor} {edited by\ \bibinfo
  {editor} {\bibfnamefont {M.}~\bibnamefont {Inguscio}}, \bibinfo {editor}
  {\bibfnamefont {W.}~\bibnamefont {Ketterle}},\ and\ \bibinfo {editor}
  {\bibfnamefont {C.}~\bibnamefont {Salomon}}}\ (\bibinfo  {publisher} {IOS
  Press},\ \bibinfo {year} {2007})\ pp.\ \bibinfo {pages}
  {289--349}\BibitemShut {NoStop}%
\bibitem [{\citenamefont {Dean}\ \emph {et~al.}(2016)\citenamefont {Dean},
  \citenamefont {Le~Doussal}, \citenamefont {Majumdar},\ and\ \citenamefont
  {Schehr}}]{dean2016}%
  \BibitemOpen
  \bibfield  {author} {\bibinfo {author} {\bibfnamefont {D.~S.}\ \bibnamefont
  {Dean}}, \bibinfo {author} {\bibfnamefont {P.}~\bibnamefont {Le~Doussal}},
  \bibinfo {author} {\bibfnamefont {S.~N.}\ \bibnamefont {Majumdar}},\ and\
  \bibinfo {author} {\bibfnamefont {G.}~\bibnamefont {Schehr}},\ }\href
  {https://doi.org/10.1103/PhysRevA.94.063622} {\bibfield  {journal} {\bibinfo
  {journal} {Physical Review A}\ }\textbf {\bibinfo {volume} {94}},\ \bibinfo
  {pages} {063622} (\bibinfo {year} {2016})}\BibitemShut {NoStop}%
\bibitem [{\citenamefont {Holten}\ \emph {et~al.}(2021)\citenamefont {Holten},
  \citenamefont {Bayha}, \citenamefont {Subramanian}, \citenamefont {Heintze},
  \citenamefont {Preiss},\ and\ \citenamefont {Jochim}}]{holten2021}%
  \BibitemOpen
  \bibfield  {author} {\bibinfo {author} {\bibfnamefont {M.}~\bibnamefont
  {Holten}}, \bibinfo {author} {\bibfnamefont {L.}~\bibnamefont {Bayha}},
  \bibinfo {author} {\bibfnamefont {K.}~\bibnamefont {Subramanian}}, \bibinfo
  {author} {\bibfnamefont {C.}~\bibnamefont {Heintze}}, \bibinfo {author}
  {\bibfnamefont {P.~M.}\ \bibnamefont {Preiss}},\ and\ \bibinfo {author}
  {\bibfnamefont {S.}~\bibnamefont {Jochim}},\ }\href
  {https://doi.org/10.1103/PhysRevLett.126.020401} {\bibfield  {journal}
  {\bibinfo  {journal} {Physical Review Letters}\ }\textbf {\bibinfo {volume}
  {126}},\ \bibinfo {pages} {020401} (\bibinfo {year} {2021})}\BibitemShut
  {NoStop}%
\bibitem [{\citenamefont {Torquato}(2018)}]{torquato2018}%
  \BibitemOpen
  \bibfield  {author} {\bibinfo {author} {\bibfnamefont {S.}~\bibnamefont
  {Torquato}},\ }\href {https://doi.org/10.1016/j.physrep.2018.03.001}
  {\bibfield  {journal} {\bibinfo  {journal} {Physics Reports}\ }\bibinfo
  {series} {Hyperuniform {{States}} of {{Matter}}},\ \textbf {\bibinfo {volume}
  {745}},\ \bibinfo {pages} {1} (\bibinfo {year} {2018})}\BibitemShut {NoStop}%
\bibitem [{\citenamefont {Dean}\ \emph {et~al.}(2019)\citenamefont {Dean},
  \citenamefont {Le~Doussal}, \citenamefont {Majumdar},\ and\ \citenamefont
  {Schehr}}]{dean2019}%
  \BibitemOpen
  \bibfield  {author} {\bibinfo {author} {\bibfnamefont {D.~S.}\ \bibnamefont
  {Dean}}, \bibinfo {author} {\bibfnamefont {P.}~\bibnamefont {Le~Doussal}},
  \bibinfo {author} {\bibfnamefont {S.~N.}\ \bibnamefont {Majumdar}},\ and\
  \bibinfo {author} {\bibfnamefont {G.}~\bibnamefont {Schehr}},\ }\href
  {https://doi.org/10.1088/1751-8121/ab098d} {\bibfield  {journal} {\bibinfo
  {journal} {Journal of Physics A: Mathematical and Theoretical}\ }\textbf
  {\bibinfo {volume} {52}},\ \bibinfo {pages} {144006} (\bibinfo {year}
  {2019})}\BibitemShut {NoStop}%
\bibitem [{\citenamefont {Macchi}(1975)}]{macchi1975}%
  \BibitemOpen
  \bibfield  {author} {\bibinfo {author} {\bibfnamefont {O.}~\bibnamefont
  {Macchi}},\ }\href {https://doi.org/10.2307/1425855} {\bibfield  {journal}
  {\bibinfo  {journal} {Advances in Applied Probability}\ }\textbf {\bibinfo
  {volume} {7}},\ \bibinfo {pages} {83} (\bibinfo {year} {1975})}\BibitemShut
  {NoStop}%
\bibitem [{\citenamefont {Johansson}(2005)}]{johansson2005}%
  \BibitemOpen
  \bibfield  {author} {\bibinfo {author} {\bibfnamefont {K.}~\bibnamefont
  {Johansson}},\ }\href {https://doi.org/10.48550/arXiv.math-ph/0510038}
  {\bibinfo {title} {Random matrices and determinantal processes}} (\bibinfo
  {year} {2005}),\ \Eprint {https://arxiv.org/abs/math-ph/0510038}
  {arXiv:math-ph/0510038} \BibitemShut {NoStop}%
\bibitem [{\citenamefont {Borodin}(2009)}]{borodin2009}%
  \BibitemOpen
  \bibfield  {author} {\bibinfo {author} {\bibfnamefont {A.}~\bibnamefont
  {Borodin}},\ }\href {https://doi.org/10.48550/ARXIV.0911.1153} {\bibinfo
  {title} {Determinantal point processes}} (\bibinfo {year} {2009})\BibitemShut
  {NoStop}%
\bibitem [{\citenamefont {Beenakker}(1997)}]{beenakker1997}%
  \BibitemOpen
  \bibfield  {author} {\bibinfo {author} {\bibfnamefont {C.~W.~J.}\
  \bibnamefont {Beenakker}},\ }\href
  {https://doi.org/10.1103/RevModPhys.69.731} {\bibfield  {journal} {\bibinfo
  {journal} {Reviews of Modern Physics}\ }\textbf {\bibinfo {volume} {69}},\
  \bibinfo {pages} {731} (\bibinfo {year} {1997})}\BibitemShut {NoStop}%
\bibitem [{\citenamefont {Potters}\ and\ \citenamefont
  {Bouchaud}(2020)}]{potters2020}%
  \BibitemOpen
  \bibfield  {author} {\bibinfo {author} {\bibfnamefont {M.}~\bibnamefont
  {Potters}}\ and\ \bibinfo {author} {\bibfnamefont {J.-P.}\ \bibnamefont
  {Bouchaud}},\ }\href {https://doi.org/10.1017/9781108768900} {\emph {\bibinfo
  {title} {A {{First Course}} in {{Random Matrix Theory}}: For {{Physicists}},
  {{Engineers}} and {{Data Scientists}}}}},\ \bibinfo {edition} {1st}\ ed.\
  (\bibinfo  {publisher} {Cambridge University Press},\ \bibinfo {year}
  {2020})\BibitemShut {NoStop}%
\bibitem [{\citenamefont {Wishart}(1928)}]{wishart1928}%
  \BibitemOpen
  \bibfield  {author} {\bibinfo {author} {\bibfnamefont {J.}~\bibnamefont
  {Wishart}},\ }\href {https://doi.org/10.2307/2331939} {\bibfield  {journal}
  {\bibinfo  {journal} {Biometrika}\ }\textbf {\bibinfo {volume} {20A}},\
  \bibinfo {pages} {32} (\bibinfo {year} {1928})},\ \Eprint
  {https://arxiv.org/abs/2331939} {2331939} \BibitemShut {NoStop}%
\bibitem [{\citenamefont {Moustakas}(2017)}]{moustakas2017}%
  \BibitemOpen
  \bibfield  {author} {\bibinfo {author} {\bibfnamefont {A.}~\bibnamefont
  {Moustakas}},\ }\href@noop {} {\emph {\bibinfo {title} {Stochastic
  {{Processes}} and {{Random Matrices}}: {{Lecture Notes}} of the {{Les Houches
  Summer School}}, {{July}} 2015}}},\ Vol.\ \bibinfo {volume} {104}\ (\bibinfo
  {publisher} {Oxford University Press},\ \bibinfo {year} {2017})\ p.\ \bibinfo
  {pages} {382}\BibitemShut {NoStop}%
\bibitem [{\citenamefont {Smith}\ \emph {et~al.}(2021)\citenamefont {Smith},
  \citenamefont {Le~Doussal}, \citenamefont {Majumdar},\ and\ \citenamefont
  {Schehr}}]{smith2021}%
  \BibitemOpen
  \bibfield  {author} {\bibinfo {author} {\bibfnamefont {N.~R.}\ \bibnamefont
  {Smith}}, \bibinfo {author} {\bibfnamefont {P.}~\bibnamefont {Le~Doussal}},
  \bibinfo {author} {\bibfnamefont {S.~N.}\ \bibnamefont {Majumdar}},\ and\
  \bibinfo {author} {\bibfnamefont {G.}~\bibnamefont {Schehr}},\ }\href
  {https://doi.org/10.1103/PhysRevE.103.L030105} {\bibfield  {journal}
  {\bibinfo  {journal} {Physical Review E}\ }\textbf {\bibinfo {volume}
  {103}},\ \bibinfo {pages} {L030105} (\bibinfo {year} {2021})}\BibitemShut
  {NoStop}%
\bibitem [{\citenamefont {Gouraud}\ \emph {et~al.}(2022)\citenamefont
  {Gouraud}, \citenamefont {Le~Doussal},\ and\ \citenamefont
  {Schehr}}]{gouraud2022}%
  \BibitemOpen
  \bibfield  {author} {\bibinfo {author} {\bibfnamefont {G.}~\bibnamefont
  {Gouraud}}, \bibinfo {author} {\bibfnamefont {P.}~\bibnamefont
  {Le~Doussal}},\ and\ \bibinfo {author} {\bibfnamefont {G.}~\bibnamefont
  {Schehr}},\ }\href {https://doi.org/10.1209/0295-5075/ac4aca} {\bibfield
  {journal} {\bibinfo  {journal} {Europhysics Letters}\ }\textbf {\bibinfo
  {volume} {137}},\ \bibinfo {pages} {50003} (\bibinfo {year}
  {2022})}\BibitemShut {NoStop}%
\bibitem [{\citenamefont {Verstraten}\ \emph {et~al.}(2025)\citenamefont
  {Verstraten}, \citenamefont {Dai}, \citenamefont {Dixmerias}, \citenamefont
  {Peaudecerf}, \citenamefont {{de Jongh}},\ and\ \citenamefont
  {Yefsah}}]{verstraten2025}%
  \BibitemOpen
  \bibfield  {author} {\bibinfo {author} {\bibfnamefont {J.}~\bibnamefont
  {Verstraten}}, \bibinfo {author} {\bibfnamefont {K.}~\bibnamefont {Dai}},
  \bibinfo {author} {\bibfnamefont {M.}~\bibnamefont {Dixmerias}}, \bibinfo
  {author} {\bibfnamefont {B.}~\bibnamefont {Peaudecerf}}, \bibinfo {author}
  {\bibfnamefont {T.}~\bibnamefont {{de Jongh}}},\ and\ \bibinfo {author}
  {\bibfnamefont {T.}~\bibnamefont {Yefsah}},\ }\href
  {https://doi.org/10.1103/PhysRevLett.134.083403} {\bibfield  {journal}
  {\bibinfo  {journal} {Physical Review Letters}\ }\textbf {\bibinfo {volume}
  {134}},\ \bibinfo {pages} {083403} (\bibinfo {year} {2025})}\BibitemShut
  {NoStop}%
\bibitem [{\citenamefont {{de Jongh}}\ \emph {et~al.}(2025)\citenamefont {{de
  Jongh}}, \citenamefont {Verstraten}, \citenamefont {Dixmerias}, \citenamefont
  {Daix}, \citenamefont {Peaudecerf},\ and\ \citenamefont
  {Yefsah}}]{dejongh2025}%
  \BibitemOpen
  \bibfield  {author} {\bibinfo {author} {\bibfnamefont {T.}~\bibnamefont {{de
  Jongh}}}, \bibinfo {author} {\bibfnamefont {J.}~\bibnamefont {Verstraten}},
  \bibinfo {author} {\bibfnamefont {M.}~\bibnamefont {Dixmerias}}, \bibinfo
  {author} {\bibfnamefont {C.}~\bibnamefont {Daix}}, \bibinfo {author}
  {\bibfnamefont {B.}~\bibnamefont {Peaudecerf}},\ and\ \bibinfo {author}
  {\bibfnamefont {T.}~\bibnamefont {Yefsah}},\ }\href
  {https://doi.org/10.1103/PhysRevLett.134.183403} {\bibfield  {journal}
  {\bibinfo  {journal} {Physical Review Letters}\ }\textbf {\bibinfo {volume}
  {134}},\ \bibinfo {pages} {183403} (\bibinfo {year} {2025})}\BibitemShut
  {NoStop}%
\bibitem [{\citenamefont {Yao}\ \emph {et~al.}(2025)\citenamefont {Yao},
  \citenamefont {Chi}, \citenamefont {Wang}, \citenamefont {Fletcher},\ and\
  \citenamefont {Zwierlein}}]{yao2025}%
  \BibitemOpen
  \bibfield  {author} {\bibinfo {author} {\bibfnamefont {R.}~\bibnamefont
  {Yao}}, \bibinfo {author} {\bibfnamefont {S.}~\bibnamefont {Chi}}, \bibinfo
  {author} {\bibfnamefont {M.}~\bibnamefont {Wang}}, \bibinfo {author}
  {\bibfnamefont {R.~J.}\ \bibnamefont {Fletcher}},\ and\ \bibinfo {author}
  {\bibfnamefont {M.}~\bibnamefont {Zwierlein}},\ }\href
  {https://doi.org/10.1103/PhysRevLett.134.183402} {\bibfield  {journal}
  {\bibinfo  {journal} {Physical Review Letters}\ }\textbf {\bibinfo {volume}
  {134}},\ \bibinfo {pages} {183402} (\bibinfo {year} {2025})}\BibitemShut
  {NoStop}%
\bibitem [{\citenamefont {Xiang}\ \emph {et~al.}(2025)\citenamefont {Xiang},
  \citenamefont {{Cruz-Col{\'o}n}}, \citenamefont {Chua}, \citenamefont
  {Milner}, \citenamefont {De~Hond}, \citenamefont {Fricke},\ and\
  \citenamefont {Ketterle}}]{xiang2025}%
  \BibitemOpen
  \bibfield  {author} {\bibinfo {author} {\bibfnamefont {J.}~\bibnamefont
  {Xiang}}, \bibinfo {author} {\bibfnamefont {E.}~\bibnamefont
  {{Cruz-Col{\'o}n}}}, \bibinfo {author} {\bibfnamefont {C.~C.}\ \bibnamefont
  {Chua}}, \bibinfo {author} {\bibfnamefont {W.~R.}\ \bibnamefont {Milner}},
  \bibinfo {author} {\bibfnamefont {J.}~\bibnamefont {De~Hond}}, \bibinfo
  {author} {\bibfnamefont {J.~F.}\ \bibnamefont {Fricke}},\ and\ \bibinfo
  {author} {\bibfnamefont {W.}~\bibnamefont {Ketterle}},\ }\href
  {https://doi.org/10.1103/PhysRevLett.134.183401} {\bibfield  {journal}
  {\bibinfo  {journal} {Physical Review Letters}\ }\textbf {\bibinfo {volume}
  {134}},\ \bibinfo {pages} {183401} (\bibinfo {year} {2025})}\BibitemShut
  {NoStop}%
\bibitem [{\citenamefont {Perrier}\ \emph {et~al.}(2019)\citenamefont
  {Perrier}, \citenamefont {Amodjee}, \citenamefont {Dussarrat}, \citenamefont
  {Dareau}, \citenamefont {Aspect}, \citenamefont {Cheneau}, \citenamefont
  {Boiron},\ and\ \citenamefont {Westbrook}}]{perrier2019}%
  \BibitemOpen
  \bibfield  {author} {\bibinfo {author} {\bibfnamefont {M.}~\bibnamefont
  {Perrier}}, \bibinfo {author} {\bibfnamefont {Z.}~\bibnamefont {Amodjee}},
  \bibinfo {author} {\bibfnamefont {P.}~\bibnamefont {Dussarrat}}, \bibinfo
  {author} {\bibfnamefont {A.}~\bibnamefont {Dareau}}, \bibinfo {author}
  {\bibfnamefont {A.}~\bibnamefont {Aspect}}, \bibinfo {author} {\bibfnamefont
  {M.}~\bibnamefont {Cheneau}}, \bibinfo {author} {\bibfnamefont
  {D.}~\bibnamefont {Boiron}},\ and\ \bibinfo {author} {\bibfnamefont
  {C.}~\bibnamefont {Westbrook}},\ }\href
  {https://doi.org/10.21468/SciPostPhys.7.1.002} {\bibfield  {journal}
  {\bibinfo  {journal} {SciPost Physics}\ }\textbf {\bibinfo {volume} {7}},\
  \bibinfo {pages} {002} (\bibinfo {year} {2019})}\BibitemShut {NoStop}%
\bibitem [{\citenamefont {Herc{\'e}}\ \emph {et~al.}(2023)\citenamefont
  {Herc{\'e}}, \citenamefont {Bureik}, \citenamefont {T{\'e}nart},
  \citenamefont {Aspect}, \citenamefont {Dareau},\ and\ \citenamefont
  {Cl{\'e}ment}}]{herce2023}%
  \BibitemOpen
  \bibfield  {author} {\bibinfo {author} {\bibfnamefont {G.}~\bibnamefont
  {Herc{\'e}}}, \bibinfo {author} {\bibfnamefont {J.-P.}\ \bibnamefont
  {Bureik}}, \bibinfo {author} {\bibfnamefont {A.}~\bibnamefont {T{\'e}nart}},
  \bibinfo {author} {\bibfnamefont {A.}~\bibnamefont {Aspect}}, \bibinfo
  {author} {\bibfnamefont {A.}~\bibnamefont {Dareau}},\ and\ \bibinfo {author}
  {\bibfnamefont {D.}~\bibnamefont {Cl{\'e}ment}},\ }\href
  {https://doi.org/10.1103/PhysRevResearch.5.L012037} {\bibfield  {journal}
  {\bibinfo  {journal} {Physical Review Research}\ }\textbf {\bibinfo {volume}
  {5}},\ \bibinfo {pages} {L012037} (\bibinfo {year} {2023})}\BibitemShut
  {NoStop}%
\bibitem [{\citenamefont {Allemand}\ \emph {et~al.}(2025)\citenamefont
  {Allemand}, \citenamefont {Dupuy}, \citenamefont {Paquiez}, \citenamefont
  {Dupuis}, \citenamefont {Ran{\c c}on}, \citenamefont {Roscilde},
  \citenamefont {Chalopin},\ and\ \citenamefont {Cl{\'e}ment}}]{allemand2025}%
  \BibitemOpen
  \bibfield  {author} {\bibinfo {author} {\bibfnamefont {M.}~\bibnamefont
  {Allemand}}, \bibinfo {author} {\bibfnamefont {G.}~\bibnamefont {Dupuy}},
  \bibinfo {author} {\bibfnamefont {P.}~\bibnamefont {Paquiez}}, \bibinfo
  {author} {\bibfnamefont {N.}~\bibnamefont {Dupuis}}, \bibinfo {author}
  {\bibfnamefont {A.}~\bibnamefont {Ran{\c c}on}}, \bibinfo {author}
  {\bibfnamefont {T.}~\bibnamefont {Roscilde}}, \bibinfo {author}
  {\bibfnamefont {T.}~\bibnamefont {Chalopin}},\ and\ \bibinfo {author}
  {\bibfnamefont {D.}~\bibnamefont {Cl{\'e}ment}},\ }\href
  {https://doi.org/10.48550/arXiv.2508.21623} {\bibinfo {title} {Observation of
  universal non-{{Gaussian}} statistics of the order parameter across a
  continuous phase transition}} (\bibinfo {year} {2025}),\ \Eprint
  {https://arxiv.org/abs/2508.21623} {arXiv:2508.21623 [cond-mat]} \BibitemShut
  {NoStop}%
\bibitem [{\citenamefont {{\"O}ttl}\ \emph {et~al.}(2005)\citenamefont
  {{\"O}ttl}, \citenamefont {Ritter}, \citenamefont {K{\"o}hl},\ and\
  \citenamefont {Esslinger}}]{ottl2005}%
  \BibitemOpen
  \bibfield  {author} {\bibinfo {author} {\bibfnamefont {A.}~\bibnamefont
  {{\"O}ttl}}, \bibinfo {author} {\bibfnamefont {S.}~\bibnamefont {Ritter}},
  \bibinfo {author} {\bibfnamefont {M.}~\bibnamefont {K{\"o}hl}},\ and\
  \bibinfo {author} {\bibfnamefont {T.}~\bibnamefont {Esslinger}},\ }\href
  {https://doi.org/10.1103/PhysRevLett.95.090404} {\bibfield  {journal}
  {\bibinfo  {journal} {Physical Review Letters}\ }\textbf {\bibinfo {volume}
  {95}},\ \bibinfo {pages} {090404} (\bibinfo {year} {2005})}\BibitemShut
  {NoStop}%
\bibitem [{\citenamefont {Frisch}\ \emph {et~al.}(2014)\citenamefont {Frisch},
  \citenamefont {Mark}, \citenamefont {Aikawa}, \citenamefont {Ferlaino},
  \citenamefont {Bohn}, \citenamefont {Makrides}, \citenamefont {Petrov},\ and\
  \citenamefont {Kotochigova}}]{frisch2014}%
  \BibitemOpen
  \bibfield  {author} {\bibinfo {author} {\bibfnamefont {A.}~\bibnamefont
  {Frisch}}, \bibinfo {author} {\bibfnamefont {M.}~\bibnamefont {Mark}},
  \bibinfo {author} {\bibfnamefont {K.}~\bibnamefont {Aikawa}}, \bibinfo
  {author} {\bibfnamefont {F.}~\bibnamefont {Ferlaino}}, \bibinfo {author}
  {\bibfnamefont {J.~L.}\ \bibnamefont {Bohn}}, \bibinfo {author}
  {\bibfnamefont {C.}~\bibnamefont {Makrides}}, \bibinfo {author}
  {\bibfnamefont {A.}~\bibnamefont {Petrov}},\ and\ \bibinfo {author}
  {\bibfnamefont {S.}~\bibnamefont {Kotochigova}},\ }\href
  {https://doi.org/10.1038/nature13137} {\bibfield  {journal} {\bibinfo
  {journal} {Nature}\ }\textbf {\bibinfo {volume} {507}},\ \bibinfo {pages}
  {475} (\bibinfo {year} {2014})}\BibitemShut {NoStop}%
\bibitem [{\citenamefont {Dixmerias}\ \emph {et~al.}(2025)\citenamefont
  {Dixmerias}, \citenamefont {Verstraten}, \citenamefont {Daix}, \citenamefont
  {Peaudecerf}, \citenamefont {de~Jongh},\ and\ \citenamefont
  {Yefsah}}]{dixmerias2025}%
  \BibitemOpen
  \bibfield  {author} {\bibinfo {author} {\bibfnamefont {M.}~\bibnamefont
  {Dixmerias}}, \bibinfo {author} {\bibfnamefont {J.}~\bibnamefont
  {Verstraten}}, \bibinfo {author} {\bibfnamefont {C.}~\bibnamefont {Daix}},
  \bibinfo {author} {\bibfnamefont {B.}~\bibnamefont {Peaudecerf}}, \bibinfo
  {author} {\bibfnamefont {T.}~\bibnamefont {de~Jongh}},\ and\ \bibinfo
  {author} {\bibfnamefont {T.}~\bibnamefont {Yefsah}},\ }\href
  {https://doi.org/10.48550/arXiv.2502.05132} {\bibinfo {title} {Fluctuation
  thermometry of an atom-resolved quantum gas: {{Beyond}} the
  fluctuation-dissipation theorem}} (\bibinfo {year} {2025}),\ \Eprint
  {https://arxiv.org/abs/2502.05132} {arXiv:2502.05132 [cond-mat]} \BibitemShut
  {NoStop}%
\bibitem [{\citenamefont {Daix}\ \emph {et~al.}(2025)\citenamefont {Daix},
  \citenamefont {Dixmerias}, \citenamefont {He}, \citenamefont {Verstraten},
  \citenamefont {de~Jongh}, \citenamefont {Peaudecerf}, \citenamefont {Zhang},\
  and\ \citenamefont {Yefsah}}]{daix2025}%
  \BibitemOpen
  \bibfield  {author} {\bibinfo {author} {\bibfnamefont {C.}~\bibnamefont
  {Daix}}, \bibinfo {author} {\bibfnamefont {M.}~\bibnamefont {Dixmerias}},
  \bibinfo {author} {\bibfnamefont {Y.-Y.}\ \bibnamefont {He}}, \bibinfo
  {author} {\bibfnamefont {J.}~\bibnamefont {Verstraten}}, \bibinfo {author}
  {\bibfnamefont {T.}~\bibnamefont {de~Jongh}}, \bibinfo {author}
  {\bibfnamefont {B.}~\bibnamefont {Peaudecerf}}, \bibinfo {author}
  {\bibfnamefont {S.}~\bibnamefont {Zhang}},\ and\ \bibinfo {author}
  {\bibfnamefont {T.}~\bibnamefont {Yefsah}},\ }\href
  {https://doi.org/10.48550/arXiv.2504.01885} {\bibinfo {title} {Observing
  {{Spatial Charge}} and {{Spin Correlations}} in a {{Strongly-Interacting
  Fermi Gas}}}} (\bibinfo {year} {2025}),\ \Eprint
  {https://arxiv.org/abs/2504.01885} {arXiv:2504.01885 [cond-mat]} \BibitemShut
  {NoStop}%
\bibitem [{\citenamefont {Eisler}(2013)}]{eisler2013}%
  \BibitemOpen
  \bibfield  {author} {\bibinfo {author} {\bibfnamefont {V.}~\bibnamefont
  {Eisler}},\ }\href {https://doi.org/10.1103/PhysRevLett.111.080402}
  {\bibfield  {journal} {\bibinfo  {journal} {Physical Review Letters}\
  }\textbf {\bibinfo {volume} {111}},\ \bibinfo {pages} {080402} (\bibinfo
  {year} {2013})}\BibitemShut {NoStop}%
\bibitem [{\citenamefont {Marino}\ \emph {et~al.}(2014)\citenamefont {Marino},
  \citenamefont {Majumdar}, \citenamefont {Schehr},\ and\ \citenamefont
  {Vivo}}]{marino2014}%
  \BibitemOpen
  \bibfield  {author} {\bibinfo {author} {\bibfnamefont {R.}~\bibnamefont
  {Marino}}, \bibinfo {author} {\bibfnamefont {S.~N.}\ \bibnamefont
  {Majumdar}}, \bibinfo {author} {\bibfnamefont {G.}~\bibnamefont {Schehr}},\
  and\ \bibinfo {author} {\bibfnamefont {P.}~\bibnamefont {Vivo}},\ }\href
  {https://doi.org/10.1103/PhysRevLett.112.254101} {\bibfield  {journal}
  {\bibinfo  {journal} {Physical Review Letters}\ }\textbf {\bibinfo {volume}
  {112}},\ \bibinfo {pages} {254101} (\bibinfo {year} {2014})}\BibitemShut
  {NoStop}%
\bibitem [{\citenamefont {{Lacroix-A-Chez-Toine}}\ \emph
  {et~al.}(2019)\citenamefont {{Lacroix-A-Chez-Toine}}, \citenamefont
  {Majumdar},\ and\ \citenamefont {Schehr}}]{lacroix-a-chez-toine2019}%
  \BibitemOpen
  \bibfield  {author} {\bibinfo {author} {\bibfnamefont {B.}~\bibnamefont
  {{Lacroix-A-Chez-Toine}}}, \bibinfo {author} {\bibfnamefont {S.~N.}\
  \bibnamefont {Majumdar}},\ and\ \bibinfo {author} {\bibfnamefont
  {G.}~\bibnamefont {Schehr}},\ }\href
  {https://doi.org/10.1103/PhysRevA.99.021602} {\bibfield  {journal} {\bibinfo
  {journal} {Physical Review A}\ }\textbf {\bibinfo {volume} {99}},\ \bibinfo
  {pages} {021602} (\bibinfo {year} {2019})}\BibitemShut {NoStop}%
\bibitem [{\citenamefont {Akemann}\ \emph {et~al.}(2022)\citenamefont
  {Akemann}, \citenamefont {Mielke},\ and\ \citenamefont
  {P{\"a}{\ss}ler}}]{akemann2022}%
  \BibitemOpen
  \bibfield  {author} {\bibinfo {author} {\bibfnamefont {G.}~\bibnamefont
  {Akemann}}, \bibinfo {author} {\bibfnamefont {A.}~\bibnamefont {Mielke}},\
  and\ \bibinfo {author} {\bibfnamefont {P.}~\bibnamefont {P{\"a}{\ss}ler}},\
  }\href {https://doi.org/10.1103/PhysRevE.106.014146} {\bibfield  {journal}
  {\bibinfo  {journal} {Physical Review E}\ }\textbf {\bibinfo {volume}
  {106}},\ \bibinfo {pages} {014146} (\bibinfo {year} {2022})}\BibitemShut
  {NoStop}%
\bibitem [{\citenamefont {Torquato}\ \emph {et~al.}(2008)\citenamefont
  {Torquato}, \citenamefont {Scardicchio},\ and\ \citenamefont
  {Zachary}}]{torquato2008}%
  \BibitemOpen
  \bibfield  {author} {\bibinfo {author} {\bibfnamefont {S.}~\bibnamefont
  {Torquato}}, \bibinfo {author} {\bibfnamefont {A.}~\bibnamefont
  {Scardicchio}},\ and\ \bibinfo {author} {\bibfnamefont {C.~E.}\ \bibnamefont
  {Zachary}},\ }\href {https://doi.org/10.1088/1742-5468/2008/11/P11019}
  {\bibfield  {journal} {\bibinfo  {journal} {Journal of Statistical Mechanics:
  Theory and Experiment}\ }\textbf {\bibinfo {volume} {2008}},\ \bibinfo
  {pages} {P11019} (\bibinfo {year} {2008})}\BibitemShut {NoStop}%
\bibitem [{\citenamefont {Jin}\ \emph {et~al.}(2024)\citenamefont {Jin},
  \citenamefont {Dai}, \citenamefont {Verstraten}, \citenamefont {Dixmerias},
  \citenamefont {Alhyder}, \citenamefont {Salomon}, \citenamefont {Peaudecerf},
  \citenamefont {{de Jongh}},\ and\ \citenamefont {Yefsah}}]{jin2024}%
  \BibitemOpen
  \bibfield  {author} {\bibinfo {author} {\bibfnamefont {S.}~\bibnamefont
  {Jin}}, \bibinfo {author} {\bibfnamefont {K.}~\bibnamefont {Dai}}, \bibinfo
  {author} {\bibfnamefont {J.}~\bibnamefont {Verstraten}}, \bibinfo {author}
  {\bibfnamefont {M.}~\bibnamefont {Dixmerias}}, \bibinfo {author}
  {\bibfnamefont {R.}~\bibnamefont {Alhyder}}, \bibinfo {author} {\bibfnamefont
  {C.}~\bibnamefont {Salomon}}, \bibinfo {author} {\bibfnamefont
  {B.}~\bibnamefont {Peaudecerf}}, \bibinfo {author} {\bibfnamefont
  {T.}~\bibnamefont {{de Jongh}}},\ and\ \bibinfo {author} {\bibfnamefont
  {T.}~\bibnamefont {Yefsah}},\ }\href
  {https://doi.org/10.1103/PhysRevResearch.6.013158} {\bibfield  {journal}
  {\bibinfo  {journal} {Physical Review Research}\ }\textbf {\bibinfo {volume}
  {6}},\ \bibinfo {pages} {013158} (\bibinfo {year} {2024})}\BibitemShut
  {NoStop}%
\bibitem [{\citenamefont {Scardicchio}\ \emph {et~al.}(2009)\citenamefont
  {Scardicchio}, \citenamefont {Zachary},\ and\ \citenamefont
  {Torquato}}]{scardicchio2009}%
  \BibitemOpen
  \bibfield  {author} {\bibinfo {author} {\bibfnamefont {A.}~\bibnamefont
  {Scardicchio}}, \bibinfo {author} {\bibfnamefont {C.~E.}\ \bibnamefont
  {Zachary}},\ and\ \bibinfo {author} {\bibfnamefont {S.}~\bibnamefont
  {Torquato}},\ }\href {https://doi.org/10.1103/PhysRevE.79.041108} {\bibfield
  {journal} {\bibinfo  {journal} {Physical Review E}\ }\textbf {\bibinfo
  {volume} {79}},\ \bibinfo {pages} {041108} (\bibinfo {year}
  {2009})}\BibitemShut {NoStop}%
\bibitem [{\citenamefont {Bornemann}(2010)}]{bornemann2010}%
  \BibitemOpen
  \bibfield  {author} {\bibinfo {author} {\bibfnamefont {F.}~\bibnamefont
  {Bornemann}},\ }\href {https://doi.org/10.1090/S0025-5718-09-02280-7}
  {\bibfield  {journal} {\bibinfo  {journal} {Mathematics of Computation}\
  }\textbf {\bibinfo {volume} {79}},\ \bibinfo {pages} {871} (\bibinfo {year}
  {2010})}\BibitemShut {NoStop}%
\bibitem [{\citenamefont {Bornemann}(2011)}]{bornemann2011}%
  \BibitemOpen
  \bibfield  {author} {\bibinfo {author} {\bibfnamefont {F.}~\bibnamefont
  {Bornemann}},\ }\href {https://doi.org/10.1007/s10208-010-9075-z} {\bibfield
  {journal} {\bibinfo  {journal} {Foundations of Computational Mathematics}\
  }\textbf {\bibinfo {volume} {11}},\ \bibinfo {pages} {1} (\bibinfo {year}
  {2011})}\BibitemShut {NoStop}%
\end{thebibliography}
\end{document}